\pdfoutput=1
\documentclass[aps,prb,reprint,showpacs,superscriptaddress]{revtex4-1}
\usepackage{graphicx}
\usepackage{graphics}
\usepackage{amsmath}
\usepackage{amssymb}
\usepackage{amsfonts}
\usepackage{dcolumn}
\usepackage{dsfont}
\usepackage{latexsym}
\usepackage{rotating}
\usepackage{color}
\usepackage{latexsym}
\usepackage{bbm}
\usepackage{subfigure}
\usepackage{float}
\usepackage{epsfig}
\usepackage{epsf}
\usepackage{psfrag}
\usepackage{bm}
\usepackage{amsthm}
\usepackage{eucal}
\usepackage{mathrsfs}
\usepackage{url}
\usepackage{braket}
\usepackage{lipsum}

\usepackage{color} 

\usepackage{hyperref}
\hypersetup{
colorlinks=true,final=true,
        linkcolor=blue,
        citecolor=blue,
        filecolor=blue,
        urlcolor=blue,
}
\begin{document}

\title{Intrinsic ferromagnetism and restrictive thermodynamic stability in MA$_2$N$_4$ and Janus VSiGeN$_4$ monolayers}

\author{Dibyendu Dey}
\thanks{Both authors contributed equally.}
\email{dibyendu.dey@maine.edu}
\affiliation{Department of Physics and Astronomy, University of Maine, Orono, Maine 04469, USA}
\author{Avijeet Ray}
\thanks{Both authors contributed equally.}
\affiliation{Department of Physics, Indian Institute of Technology Roorkee, Roorkee, Uttarakhand 247667, India}
\author{Liping Yu}
\email{liping.yu@maine.edu}
\affiliation{Department of Physics and Astronomy, University of Maine, Orono, Maine 04469, USA}

\begin{abstract}
The seminal experimental discovery of the remarkably stable MoSi$_2$N$_4$ monolayer has led to a handful of predicted magnetic two-dimensional (2D) materials in the MA$_2$Z$_4$ family (M = transition metals, A = Si, Ge, and Z = N, P, As). These magnetic monolayers were predicted to be dynamically stable, but none of them has been synthesized to date. In this Research Letter, from first-principles thermodynamic stability analysis, we demonstrate that only the nitrides are thermodynamically stable and this occurs under N-rich conditions. Based on this finding, we propose two ferromagnetic, semiconducting Janus monolayers in the family: VSiGeN$_4$ and VSiSnN$_4$. They are both dynamically and thermally stable, but only the former is thermodynamically stable. Intriguingly, Janus VSiGeN$_4$ and VSiSnN$_4$ monolayers show weak in-plane anisotropy compared with the VSi$_2$N$_4$ monolayer. These two emerging Janus magnetic semiconductors offer opportunities for studying 2D magnetism and spin control for spintronics applications.
\end{abstract}

\maketitle

Atomically thin magnetic monolayers provide the ideal platform to study magnetism and spintronics device concepts in the two-dimensional (2D) limit, where magnetic properties can be effectively controlled or switched by proximity effects and external perturbations such as the magnetic field, electric field, defects, strain, optical doping, etc.~\cite{gate_tunable_FGT, 2D_hetero_mag, electric_control_Cri3, electric_switch, defects_VSe2, surface_vacancy_magnetism, magnetostriction, optical_defect_RuCl3} In the last few years, a large number of 2D intrinsic magnetic materials have been experimentally discovered and theoretically proposed.~\cite{2d_review_Blei, 2d_review_Jiang} The majority can be classified into three broad groups, namely, transition metal halides,~\cite{CrI3_Huang, CrBr3_Zhang} transition metal chalcogenides~\cite{gate_tunable_FGT, FePS3_Lee}, and MXenes and MXene analogs~\cite{MXene_carbide, MXene_nitride}.  These materials have exhibited a wide spectrum of magnetic and electronic properties. However, they also carry various disadvantages for practical manufacturing applications. For example, halides and MXenes mostly are vulnerable and reactive in the presence of ambient air and water~\cite{CrI3_deg_Sch, mxene_deg_Zhang}. Besides, experimentally reported Curie temperatures $T_{C}$  are usually much lower than the room temperature \cite{CrI3_Huang, CrBr3_Zhang, gate_tunable_FGT}. Hence, the search for new stable magnetic 2D materials continues to be a major research direction in this field. 

An emerging group of 2D materials are septuple atomic layers of ternary transition metal pnictides in the form of MA$_{2}$Z$_{4}$, where M is the transition metal, A = Si or Ge, and Z = N, P, or As~\cite{MA2Z4_wang}. The first member of this group discovered is the semiconductor MoSi$_{2}$N$_{4}$ centimeter-scale monolayer, which was successfully synthesized in 2020~\cite{MSi2N4_science}.
This material exhibited exceptional stability to air, water, acid, and heat. Following this discovery, first-principles calculations found tens of dynamically stable MA$_{2}$Z$_{4}$ compounds~\cite{MA2Z4_wang, Ding_JPCC} with properties of interest for various applications such as spintronics ~\cite{spin-valley_VSi2N4_Cui, valley_VSi2N4_Feng}, superconductors~\cite{superconductor_MSi2N4_Yan}, and catalysts~\cite{Catalyst_Liu, Catalyst_Chen}.  Of these, nine were predicted to be magnetic, including five nitrides (VSi$_{2}$N$_{4}$, NbSi$_{2}$N$_{4}$, NbGe$_{2}$N$_{4}$, TaGe$_{2}$N$_{4}$, and YSi$_{2}$N$_{4}$), two phosphides (VSi$_{2}$P$_{4}$ and  VGe$_{2}$P$_{4}$), and two arsenides (VSi$_{2}$As$_{4}$ and VGe$_{2}$As$_{4}$)~\cite{MA2Z4_wang, Ding_JPCC}. Remarkably, they have all been determined to be ferromagnetic and semiconducting except YSi$_{2}$N$_{4}$, which is also ferromagnetic but is metallic~\cite{Ding_JPCC}.

However, these 2D ferromagnetic MA$_{2}$Z$_{4}$ materials have yet to be synthesized. They were mainly predicted based on their dynamical stability,
\emph{ab initio} molecular dynamics, and negative heat formation energies (meaning energetically more favorable than their constituent elemental phases). The thermodynamic stability with respect to their stable competing phases has yet to be determined. It is well known that those dynamically stable compounds can often be thermodynamically unstable and be easily decomposed into their competing phases. Such thermodynamic instability can make their experimental synthesis very challenging especially under thermodynamic (quasi)equilibrium conditions. 

Above or near room temperature long-range magnetic ordering is necessary to realize spintronics functionality in 2D materials. Unfortunately, none of these proposed 2D ferromagnetic MA$_{2}$Z$_{4}$ compounds has been identified to possess an intrinsic long-range magnetic order with a $T_{C}$ above the room temperature. For example, density functional theory (DFT) and Monte Carlo calculations suggested that monolayers VSi$_2$N$_4$ and VSi$_2$P$_4$ in the same structure as MoSi$_{2}$N$_{4}$ have magnetic phase transition temperatures above room temperature, but they are XY magnets.~\cite{valley_VSi2N4_Feng, Lake_APL} Nevertheless, the VSi$_{2}$P$_{4}$ monolayer in a different crystal structure is the only reported 2D magnet with an out-of-plane easy axis, but its predicted $T_{C}$ is merely about 90 K~\cite{MA2Z4_wang}.

\begin{figure}
\includegraphics[width=0.6\columnwidth]{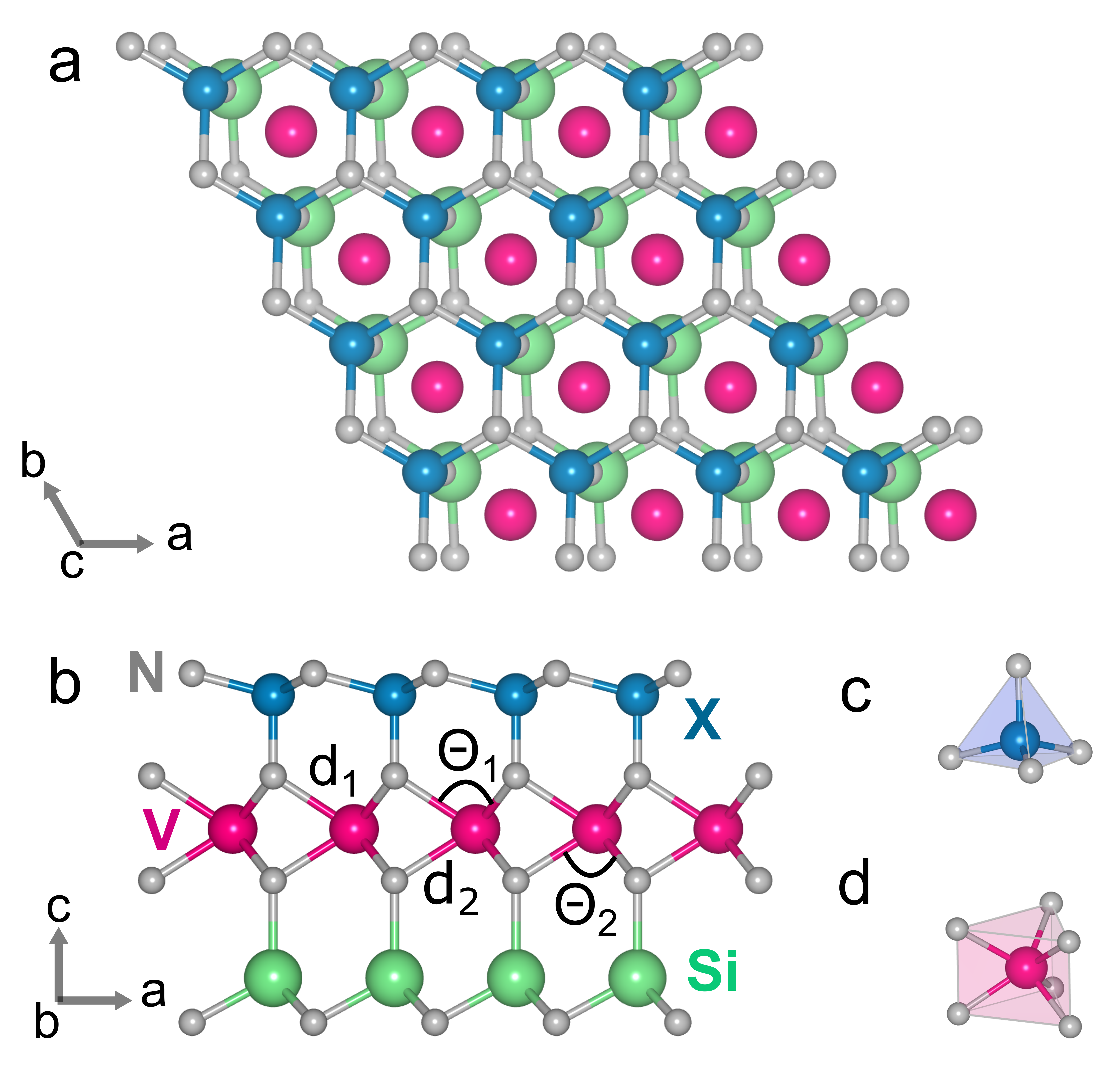}
\caption{(a) Top and (b) side views of the crystal structures of ternary layered VSiXN$_{4}$ (X = Ge, Sn) Janus monolayers. V atoms are depicted in pink, X atoms are in blue, Si atoms are in green, and N atoms are in gray. For a VSi$_2$N$_4$ monolayer, X is replaced by Si. N-V-N bond angles ($\sim 90^{\circ}$) and V-N bond lengths for the top (bottom) layer are indicated as $\theta_1$ ($\theta_2$) and d$_1$ (d$_2$), respectively. The crystal field environments: (c) tetrahedral for Si/X and (d) trigonal prismatic for V atoms.}
\label{fig1}
\end{figure}

In this Research Letter, from first-principles thermodynamic stability analysis (cf. computational methods in Supplemental Material Sec. S1) of all the nine 2D magnetic MA$_{2}$Z$_{4}$ monolayers, we explain why they have not been synthesized so far and suggest possible routes to grow them. Two-dimensional Janus crystals and their superlattices are a new class of 2D materials that offer extraordinary physical, chemical, and quantum properties~\cite{Janus_Yag,Janus_Yarjovi}. Although the electronic structures and piezoelectric properties of two non magnetic Janus MSiGeN$_4$ (M = Mo and W) monolayers were studied in a recent work~\cite{Guo_Janus}, no magnetic Janus monolayers within the same family have been predicted to date.  Here, we propose two  ferromagnetic Janus monolayers, namely, VSiGeN$_{4}$ and VSiSnN$_{4}$, which can be viewed as replacing one of the Si layers in a VSi$_{2}$N$_{4}$ monolayer with Ge or Sn  (Fig.~\ref{fig1}). We demonstrate that thermodynamically stable VSiGeN$_{4}$ and dynamically stable VSiSnN$_{4}$ monolayers are weak XY-type 2D ferromagnets with transition temperatures above 300 K, and both of them are small-gap semiconductors.

Thermodynamic stability is a key property that must be addressed to predict new materials  and understand their experimental synthesizability. It refers to the relative stability with respect to all the possible competing phases in thermodynamic equilibrium. For a material to be thermodynamically stable during growth, the chemical potentials of its constituent elements must satisfy a set of conditions. Taking MoSi$_2$N$_4$ as an example, one condition is that \begin{equation}
\Delta\mu_\textrm{Mo}+2\Delta\mu_\text{Si}+4\Delta\mu_\text{N}=\Delta H_f(\text{MoSi}_2\text{N}_4), 
\label{eqn1}
\end{equation}
where $\Delta H_f(\text{MoSi}_2\text{N}_4)$ is the enthalpy of formation, which can be calculated from first-principles. $\Delta\mu_\textrm{Mo}$, $\Delta\mu_\text{Si}$, and $\Delta\mu_\text{N}$ are the relative chemical potentials of elemental Mo, Si, and N with respect to their solid bulk components (extreme rich limits), respectively. These chemical potentials depend on the experimental growth conditions and are regarded as variable in the formalism. They are further bound by (i) the values that will cause precipitation of solid elemental Mo, Si, and N, so that 
\begin{align}
\Delta\mu_\text{Mo} \leq 0,\; \Delta\mu_\text{Si} \leq 0, \; \Delta\mu_\text{N} \leq 0,
\label{eqn2}
\end{align}
and (ii) by the values that will lead to the precipitation of competing binary phases MoN and Si$_3$N$_4$, so that 
\begin{align}
\Delta\mu_\text{Mo}+\Delta\mu_\text{N} < \Delta H_f\text{(MoN)},\\
3\Delta\mu_\text{Si}+4\Delta\mu_\text{N}< \Delta H_f(\text{Si}_3\text{N}_4).
\label{eqn3}
\end{align}
The ranges of $\Delta\mu_\textrm{Mo}$, $\Delta\mu_\text{Si}$, and $\Delta\mu_\text{N}$ that satisfy Eqs.(\ref{eqn1})$-$(\ref{eqn3}) define the thermodynamic stability region, where possible equilibrium experimental growth conditions are suggested. No solution to these equations means that the material investigated is thermodynamically unstable and its equilibrium experimental synthesis would be quite challenging or impossible as kinetic barriers between different phases at high temperatures are often ignorable or small.

\begin{figure}
\includegraphics[width=0.9\columnwidth]{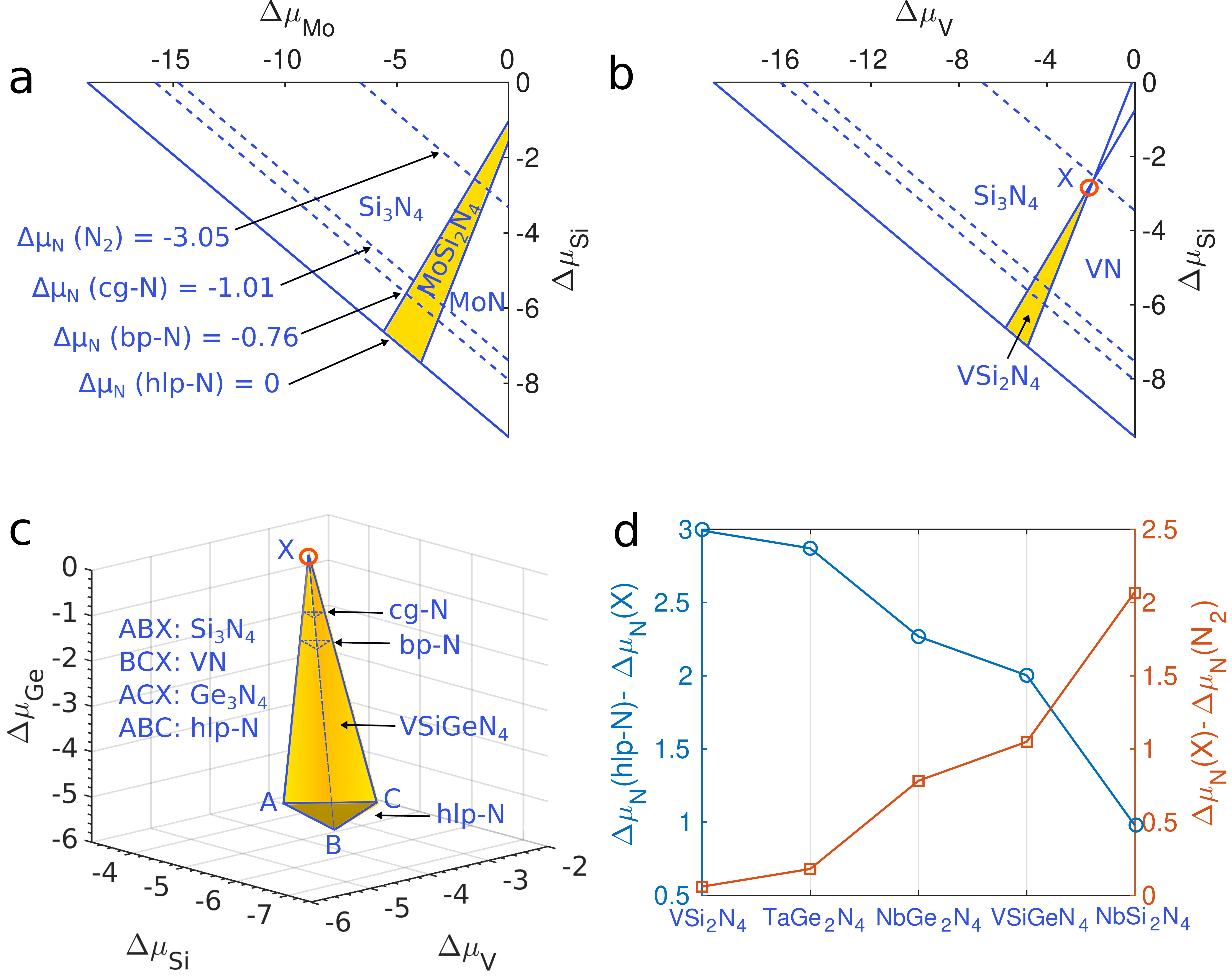}
\caption{2D or 3D chemical potential regions (shaded in yellow) where (a) MoSi$_2$N$_4$, (b) VSi$_2$N$_4$, and (c) VSiGeN$_4$ monolayers are thermodynamically stable against the formation of constituent elements and competing binary compounds. The shaded area inside the solid blue line (dashed blue line) in the phase diagram of MoSi$_2$N$_4$ is the chemical potential region where the monolayer can be synthesized using chemically rich (poor) N. X (marked with a red circle) represents the vertex of the chemical potential region, up to which the respective compounds can be synthesized using chemically poor N. (d) The difference in chemical potential of N at X and N$_2$ gas (hlp-N) is shown on the right (left) axes. All the energies are in eV.}
\label{fig2}
\end{figure}
As a test, we first perform the above thermodynamic stability analysis for two nonmagnetic monolayers MoSi$_2$N$_4$ and WSi$_2$N$_4$, which have been experimentally synthesized~\cite{MSi2N4_science}. Consistent with experiment, our calculated results shown in Fig.~\ref{fig2}(a) and Supplemental Material Fig. S2(a) demonstrate that these two compounds are indeed thermodynamically stable. Their stability regions are highlighted in yellow in the plots. It is worth mentioning that this is defined relative to the chemical potential of hexagonal layered polymeric nitrogen (hlp-N), based on which $\Delta H_f$ are calculated. This hlp-N phase represents the extreme N-rich growth condition and was experimentally synthesized near 250 GPa~\cite{hlp-N}. The N$_2$ gas under normal conditions is about 3.05 eV lower in chemical potential (poorer) than hlp-N; in between, there are another two metastable phases, cubic gauche polymeric nitrogen (cg-N)~\cite{cg-N} and a polymeric nitrogen allotrope with  black phosphorus structure (bp-N)~\cite{bp-N}, which were also experimentally synthesized.  As displayed in Fig.~\ref{fig2}(a) and Supplemental Material Fig. S2(a), the thermodynamic stability regions of MoSi$_2$N$_4$ and WSi$_2$N$_4$ both cover from extreme N-rich conditions to  conditions that are much poorer than the N$_2$ gas. 

We now perform the same thermodynamic stability analysis for all nine magnetic MA$_2$Z$_4$ monolayers. Computational details are given in Supplemental Material Sec. S1. Our calculated results are shown in Fig.~\ref{fig2}(b) and Supplemental Material Figs. S2(b)-S2(d). Although these monolayers were all predicted to be dynamically stable and have a negative formation enthalpy, our calculations reveal that only nitrides, except YSi$_{2}$N$_{4}$ are thermodynamically stable and this occurs only under some conditions for N that are richer than N$_2$ gas. Their stability regions are all much narrower than those for MoSi$_2$N$_4$ and WSi$_2$N$_4$. Furthermore, to understand the relative stability of magnetic MA$_2$N$_4$, the distance between the vertex of each chemical potential region (marked as X and shown with a red circle) and N$_2$ gas has been calculated. $\Delta\mu_N$(X)$-$ $\Delta\mu_N$(N$_2$) increases in the following order: VSi$_2$N$_4$ $\rightarrow$ \text{TaGe}$_{2}$N$_{4}$ $\rightarrow$ \text{NbGe}$_{2}$N$_{4}$ $\rightarrow$ \text{NbSi}$_{2}$N$_{4}$, along which the respective regions are far from the chemical potential of N$_2$ gas [Fig.~\ref{fig2}(d)]. This means that VSi$_2$N$_4$ is the most stable among these magnetic nitrides. These results suggest that the experimental growth of these nitrides under thermodynamic equilibrium conditions will be difficult as extreme N-rich conditions are necessary. VSi$_2$N$_4$ may be grown via the reaction
\begin{align}
2\text{Si}_3\text{N}_4 + 3\text{VN} + \text{N (solid)} \longrightarrow 3\text{VSi}_2\text{N}_4,
\label{eqn2}
\end{align}
where the solid N reactant can be hlp-N, cg-N, or bp-N.  For phosphides (VSi$_{2}$P$_{4}$ and  VGe$_{2}$P$_{4}$) and arsenides (VSi$_{2}$As$_{4}$ and VGe$_{2}$As$_{4}$), we cannot find any thermodynamically stable region. Thus, under equilibrium growth conditions, the competing binary phases (e.g., V$_3$P, VP, SiP$_2$, GeP$_3$, etc., for phosphides and V$_3$As, VAs, SiAs$_2$, GeAs$_2$, etc., for arsenides) of these materials would form rather than the materials themselves.

To identify more stable intrinsic ferromagnetic 2D MA$_2$Z$_4$ compounds, we extend our first-principles thermodynamic stability analysis to two  Janus compounds, VSiGeN$_4$ and VSiSnN$_4$. They are derived from the VSi$_2$N$_4$ monolayer, which has been demonstrated to be thermodynamically stable. Janus monolayers VSiXN$_{4}$ (X = Ge, Sn) can be viewed as a VN$_2$ layer sandwiched between Si-N and X-N layers (Fig.~\ref{fig1}). Due to the presence of different group-IV elements in the top and bottom layers, such Janus monolayers lack both inversion and out-of-plane mirror symmetries in contrast to the VSi$_2$N$_4$ monolayer. 

Our thermodynamic stability study reveals that both VSiGeN$_{4}$ and VSiSnN$_{4}$ monolayers have negative formation enthalpies (using hlp-N) of $-$4.329 eV and $-$2.132 eV, respectively. However, only monolayer VSiGeN$_{4}$ is found to be thermodynamically stable. Figure~\ref{fig2}(c) shows our calculated stability region for VSiGeN$_{4}$ and its competing phases Si$_3$N$_4$, Ge$_3$N$_4$, and VN. As expected, this thermodynamic stability region is smaller than that of VSi$_2$N$_4$ and stays ~1 eV farther from the chemical potential of N$_2$ gas [Fig.~\ref{fig2}(d)]. However, advantageously, this stability region also covers the chemical potentials of both cg-N and bp-N phases. The VSiGeN$_{4}$ monolayer may thus be synthesized via 
\begin{align}
\text{Si}_3\text{N}_4 + \text{Ge}_3\text{N}_4 + 3\text{VN} + \text{cg-N} \longrightarrow 3\text{VSiGeN}_4.
\label{eqn2}
\end{align}
For the VSiSnN$_4$ monolayer, no thermodynamic stability region is found in this paper.

\begin{figure}
\includegraphics[width=0.9\columnwidth]{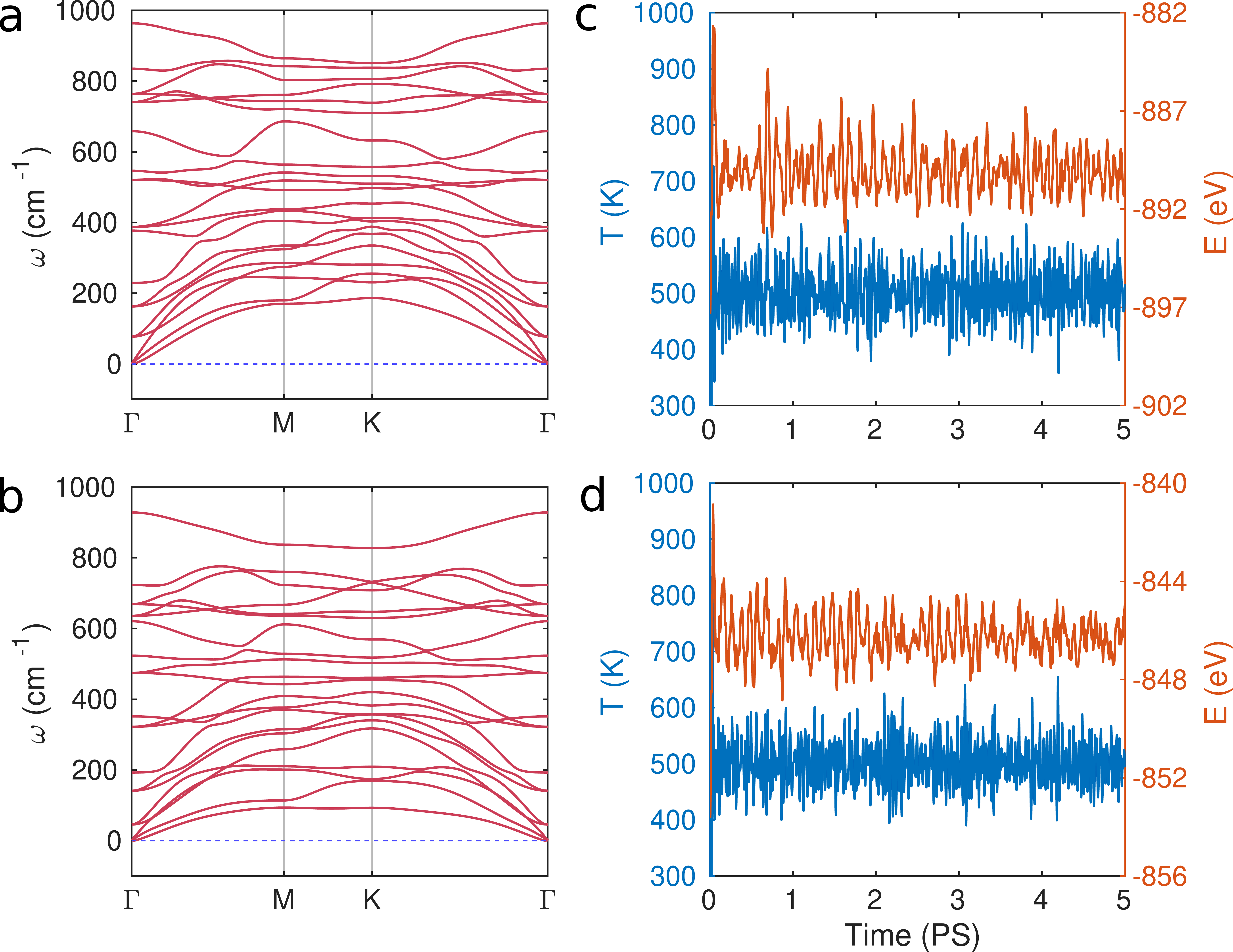}
\caption{(a) and (b) Phonon band structures, and (c) and (d) temperatures and total energy fluctuations after 5 ps of \emph{ab initio} molecular dynamics simulations at 500 K of VSiGeN$_{4}$ (top panels) and VSiSnN$_{4}$ (bottom panels) Janus monolayers.}
\label{fig3}
\end{figure}
Both Janus monolayers VSiGeN$_{4}$ and VSiSnN$_{4}$ are also found to be dynamically and thermally stable. Figures~\ref{fig3}(a) and \ref{fig3}(b) show our calculated phonon spectra along the high-symmetry directions of the hexagonal Brillouin zone (BZ). No negative frequencies are found in either case. Figures~\ref{fig3}(c) and \ref{fig3}(d) show our \emph{ab initio} molecular dynamics simulation results for both Janus monolayers at 500 K. No indication of bond breaking or significant structural distortions is seen after 5 ps of simulation time (1 fs time step). Such stability features render these Janus materials also synthesizable using non-equilibrium techniques.~\cite{Lu_Janus,Q_adv_matr}

\begin{figure}
\includegraphics[width=0.7\columnwidth]{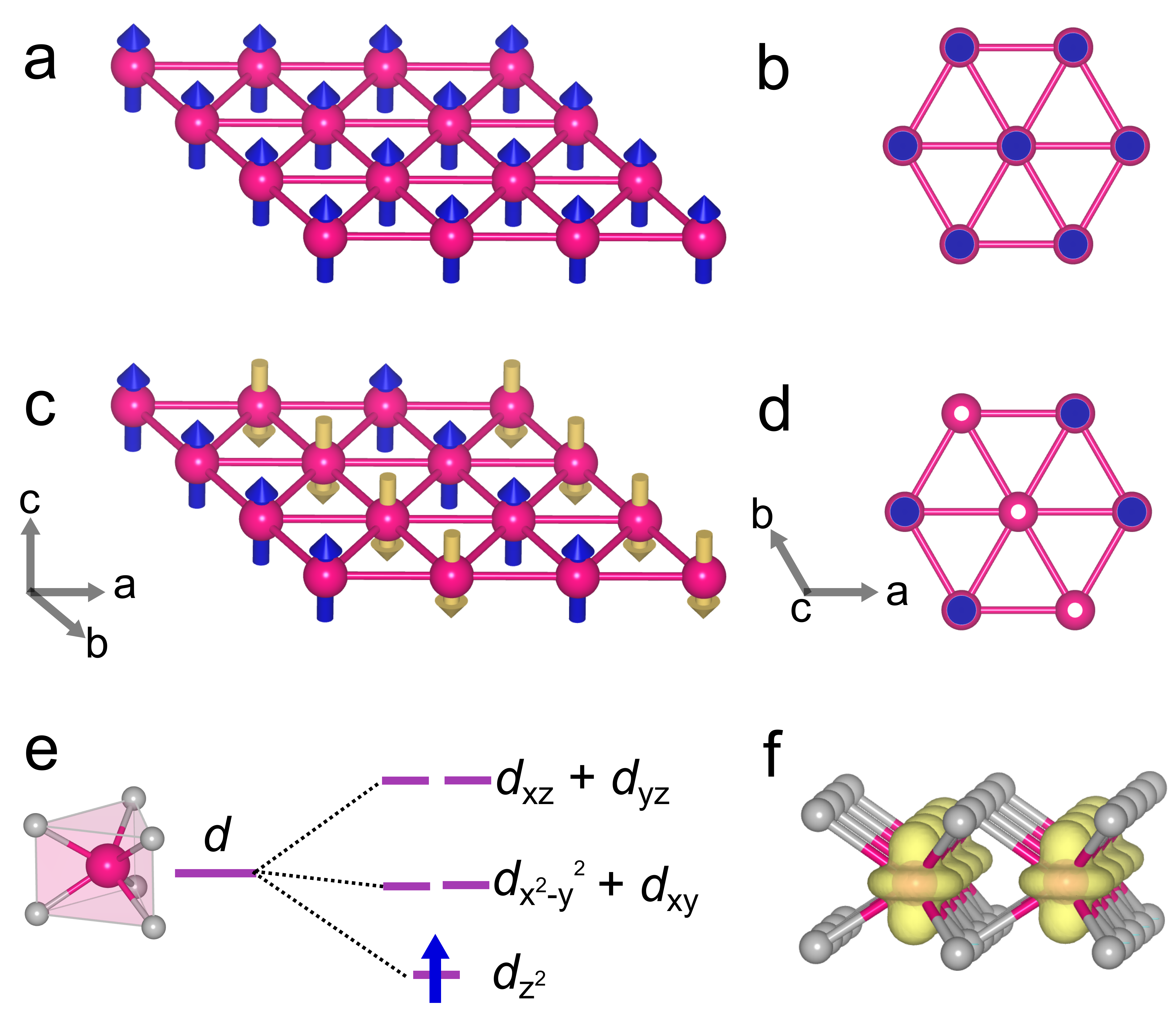}
\caption{Side and top views of the effective triangular lattice formed by V ions in VSi$_2$N$_4$ and VSiXN$_{4}$ Janus monolayers showing (a) and (b) FM and (c) and (d) stripe AFM states. V ions are shown as pink spheres. Up and down spins are indicated by blue and gold arrows, respectively. (e) A schematic of the energy levels of $d$ orbitals in a trigonal prismatic crystal field environment. Here, only the middle layer containing V ions is displayed since the non-magnetic top and bottom layers of these monolayers do not contribute to the spin density distribution. The single valence electron of the V$^{4+}$ ion predominately occupies the $d_{z^2}$ orbital, and the corresponding spin density distribution is shown in (f).}
\label{fig4}
\end{figure}
Now, we turn to study the magnetic properties of VSiGeN$_{4}$ and VSiSnN$_{4}$ monolayers. To identify the magnetic ground state, we calculate the total energy difference between the ferromagnetic (FM) and stripe antiferromagnetic (AFM) spin configurations as depicted in Figs.~\ref{fig4}(a)$-$\ref{fig4}(d) using the generalized gradient approximation (GGA)+$U$ method~\cite{dudarev}. To include correlation effects in V 3$d$ orbitals, we choose $U$ = 3.2~eV, at which the GGA+$U$ calculated energy difference between the AFM and FM states matches with the revised Heyd-Scuseria-Ernzerhof screened hybrid functional (HSE06)~\cite{hse06} calculated one. (cf. Supplemental Material Fig. S3). The $U$ value determined in this way has successfully predicted correct magnetic, structural properties in numerous systems~\cite{benchU1, benchU2}.

Our GGA+$U$ calculations find that both Janus monolayers stabilize into a FM ground state, like the VSi$_2$N$_4$ monolayer. As one can see from Table~\ref{table:1}, the FM states in VSiGeN$_{4}$ and VSiSnN$_{4}$ are 65.5 and 44.8~meV lower in energy than their AFM states, respectively. Magnetic moments $m_{V}$ obtained from our GGA+$U$ calculations agree qualitatively with the ionic description that gives a V$^{4+}$: 3$d^{1}$ electronic configuration. In a trigonal prismatic crystal field environment, V $d$ orbitals split into the low-lying $d_{z^2}$ orbital followed by the ($d_{xy}$+$d_{x^2-y^2}$) and ($d_{xz}$+$d_{yz}$) orbitals   [Fig.~\ref{fig4}(e)], respectively. Our calculated spin density distribution of the occupied V $d$ orbital [Fig.~\ref{fig4}(f)] shows that the top valence band in the majority-spin channel is predominantly of $d_{z^2}$ orbital character. Here, the V-N-V angle is close to 90$^\circ$, and within this orthogonal setup, superexchange interaction between the V atoms mediated by the neighboring N atoms is the dominating exchange mechanism for ferromagnetism. This is in agreement with the Goodenough-Kanamori rules.~\cite{goodenough, kanamori} As a result, VSi$_2$N$_4$ and VSiXN$_4$ monolayers adopt a FM ordered ground state.

\begin{table}
\centering
\begin{center}
\begin{tabular}{p{1.3cm} p{.9cm} p{1.4cm} p{1.0cm} p{1.0cm} p{1.0cm} p{1.0cm} p{0.01cm}}
\hline 

\centering Monolayer &\centering $a$ & \centering $E_g$ & \centering $m_{V}$ & \centering $\Delta E_m$ & \centering MAE & \centering $T_{BKT}$ & \\
 &  \centering (\AA) & \centering (eV) & \centering ($\mu_{B}$) & \centering (meV)& \centering ($\mu$eV) & \centering (K) & \\
 \hline
 \centering VSi$_2$N$_{4}$  & \centering 2.90 & \centering 0.50 (D) & \centering 1.10 & \centering 88.7 &\centering -59 & \centering  687 & \\

 \hline
 \centering VSiGeN$_{4}$  & \centering 2.97 & \centering 0.48 (D) &  \centering 1.11 & \centering 65.5 &\centering -20 & \centering 507 & \\

 \hline
 \centering VSiSnN$_{4}$ & \centering 3.07 & \centering 0.85 (ID) & \centering 1.14 & \centering 44.8 & \centering -10 & \centering  347 &\\

\hline         
\hline 
\end{tabular}
\caption{Lattice constants $a$, band gap ($E_g$), V magnetic moments ($m_V$), energy differences between the AFM and FM spin configurations [$\Delta$E$_{m}$ = (E$_{AFM}$ $-$ E$_{FM}$)/f.u.], magnetic anisotropy energy (MAE), which includes both magnetocrystalline anisotropy (MCA) and magnetic shape anisotropy (MSA) and transition temperatures $T_{BKT}$ of VSi$_2$N$_4$ and VSiXN$_4$ monolayers. The band gap $E_g$ values mentioned in the third column are obtained using HSE06 and agree qualitatively with the band gap values obtained within the GGA+$U$ method:  VSi$_2$N$_4$, 0.3 eV [direct semiconductor (D)], VSiGeN$_4$, 0.15  eV (D), VSiSnN$_4$: 0.28 eV [indirect semiconductor (ID)].}
\label{table:1}
\end{center}
\end{table}

 According to the Mermin-Wagner theorem~\cite{mermin}, long-range magnetic order should be absent in the 2D isotropic spin systems at any finite temperature. However, magnetic anisotropy  removes this restriction and stabilizes magnetic order in 2D materials~\cite{CrI3_Huang,PNAS_CrX3,Cai_CrCl3,Liu_CoPS3}.
 To determine the nature of magnetic anisotropy, we first calculate magnetocrystalline anisotropy (MCA) using the GGA+$U$ method including spin-orbit coupling (SOC). MCA is defined as the energy difference between two FM spin configurations, where magnetic moments are directed along the $x$ and $z$ directions, respectively. For the VSiGeN$_4$ monolayer, we obtain a negative MCA ($-$4$~\mu$eV) and a positive MCA (4$~\mu$eV) for the VSiSnN$_4$ monolayer. Since MCA is significantly small here, magnetic shape anisotropy (MSA), which originated from dipole-dipole interactions may become relevant and contribute to the magnetic anisotropy energy (MAE). The calculated MSA values for VSiGeN$_4$ ($-$16$~\mu$eV) and VSiSnN$_4$ ($-$14$~\mu$eV) monolayers are found to be greater than the magnitude of MCA values, whereas for VSi$_2$N$_4$, MCA ($-$42$~\mu$eV) remains higher than MSA ($-$17$~\mu$eV) (cf. Supplemental Material Sec. S5 for details). As a result, after the inclusion of MSA, the total MAE for Janus monolayers becomes negative (easy-plane anisotropy), and owing to its small magnitude (Table~\ref{table:1}), VSiXN$_4$ monolayers can be regarded as weak 2D-XY magnets like the CrCl$_3$ monolayer.~\cite{PNAS_CrX3,Akram_NL_CrX3} A quasi-long-range order with a divergent correlation length could be observed below the Berezinskii-Kosterlitz-Thouless (BKT)~\cite{BKT} transition temperature for these planar magnets. However, due to the weak anisotropy, in-plane magnetization in both Janus monolayers could easily be tuned to out-of-plane long-range order by perturbations such as strains and defects. 

 The magnetic transition temperature $T_{BKT} =$ 1.335 $J/k_B$ is obtained from the results of the Monte Carlo simulation for 2D-XY magnets on a triangular lattice~\cite{BKT_Zhang_triangular} (cf. Supplemental Material Sec. S4). We find that the transition temperatures of VSiGeN$_4$ (507 K) and VSiSnN$_4$ (347 K) are lower than that of VSi$_2$N$_4$ (687 K), but they are still higher than the room temperature.

\begin{figure}
\includegraphics[width=0.80\columnwidth]{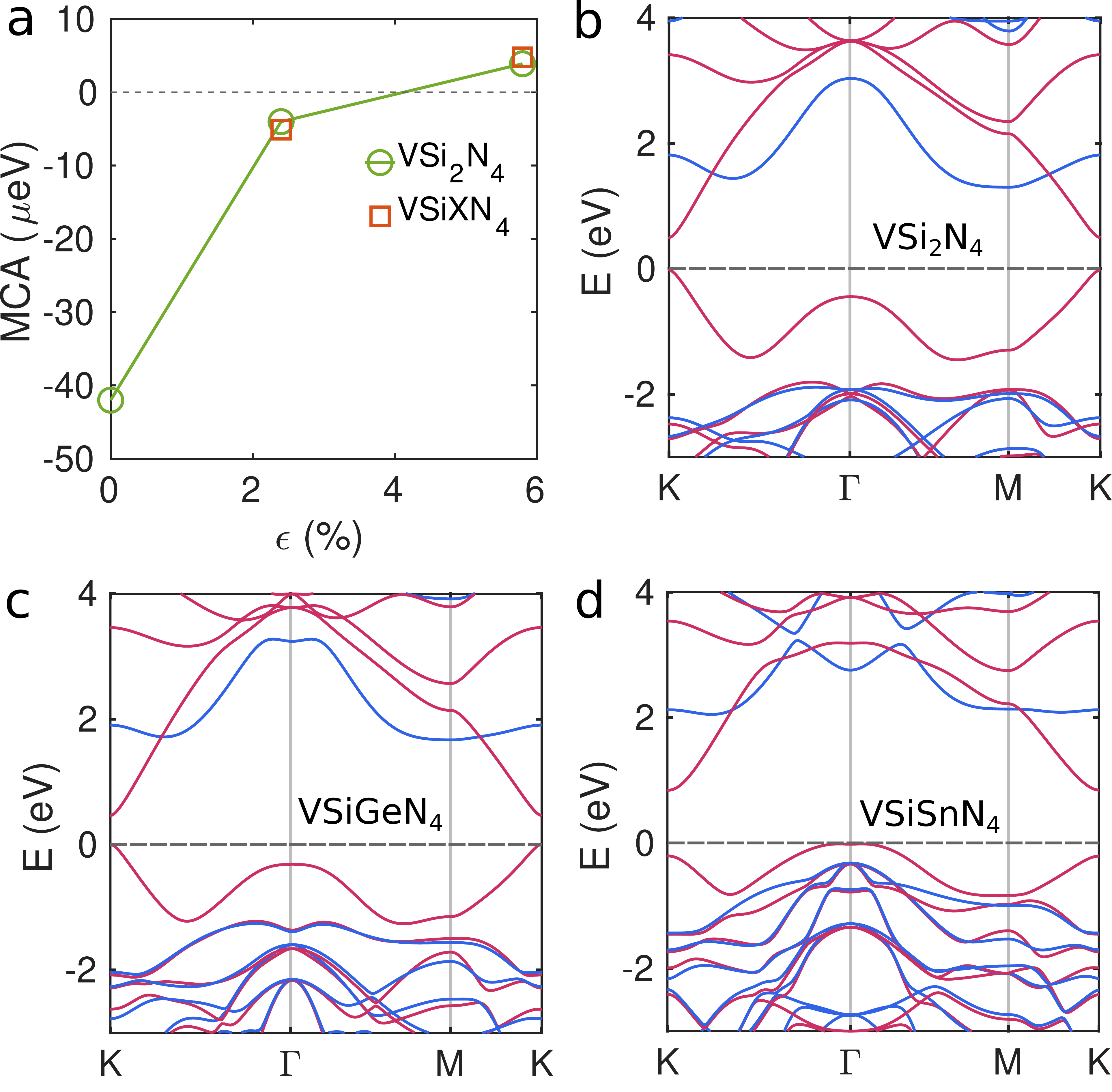}
\caption{(a) Comparison of magnetocrystalline anisotropy (MCA) values between strained monolayers VSi$_2$N$_4$ and Janus monolayers VSiXN$_4$ (X = Ge, Sn). $\epsilon$ is the uniform biaxial strain with respect to the VSi$_2$N$_4$ optimized lattice constant. Band structures of (b) VSi$_2$N$_{4}$, (c) VSiGeN$_{4}$, and (d) VSiSnN$_{4}$ monolayers within HSE06 method in the FM ground state. Red and blue lines in the band structures represent the majority and minority spin channels, respectively.}
\label{fig5}
\end{figure}
To understand the nature of MCA variations in Janus monolayers, we also calculated the MCA for strained VSi$_2$N$_4$ monolayers.  Due to the presence of larger-size atoms, VSiGeN$_4$ and VSiSnN$_4$ Janus structures have  lattice constants that are 2.4 and 5.8\% larger than that of monolayer VSi$_2$N$_4$  (Table~\ref{table:1}), respectively. Figure~\ref{fig5}(a) shows a comparison of MCA values between strained monolayer VSi$_2$N$_4$ and Janus monolayers. Clearly, one can observe that the MCA values of Janus monolayers (marked as red squares) are almost identical to the MCA values of VSi$_2$N$_4$ monolayers at the same lattice constants of corresponding Janus monolayers. This means that the key factor that drives the transition from the in-plane MCA in VSi$_2$N$_4$ and VSiGeN$_4$ to the out-of-plane MCA in the VSiSnN$_4$ monolayer is the biaxial in-plane tensile strain. The strain effects of MCA in VSi$_2$N$_4$  have already been theoretically studied in the past~\cite{spin-valley_VSi2N4_Cui}. These reported results, agreeing well with ours, indicate that switching the MCA from in-plane to out-of-plane in VSi$_2$N$_4$ needs more than 4\% of biaxial strains, which would be very challenging for experimental realization and disadvantageous for device applications. Here our results demonstrate that Janus engineering is a useful (and probably more viable) method to tune MCA and other properties of interest.

In addition to remarkable stability and magnetic properties, these two Janus monolayers exhibit excellent electronic properties. Figures~\ref{fig5}(b)$-$\ref{fig5}(d) depict the electronic band structures of VSi$_2$N$_4$ and VSiXN$_4$ monolayers as obtained from HSE06. VSi$_2$N$_4$ and VSiGeN$_4$ monolayers are direct-gap semiconductors with band gap $E_g$ $\sim$ 0.5 eV. Interestingly, the nature of the band gap changes from direct to indirect when $E_g$ increases to 0.85 eV for VSiSnN$_4$. This band gap switching and these different $E_g$ values make VSiXN$_4$ monolayers suitable candidates to operate at variable voltages, frequencies, and temperature ranges for device applications. Note that these band structures are consistent with those determined within the GGA+$U$ method (Supplemental Material Fig. S7). 

Finally, we shed light on the valley polarization, which is a measure of spin-valley coupling. This quantity can be estimated from the energy difference $\Delta E_{sv}$ between valance band maxima or conduction band minima at the $-K$ and $+K$ points. For this purpose, we calculate the band structures within GGA+$U$+SOC (see Supplemental Material Fig. S8) with the magnetization direction along +$x$ and +$z$ directions, respectively. A relatively large valley polarization of 64 meV can be induced in the VSi$_2$N$_4$ monolayer by changing the spin directions from +$x$ to +$z$, consistent with a previous report~\cite{spin-valley_VSi2N4_Cui}. On the other hand, the valley polarization can be induced in VSiXN$_{4}$ monolayers through the valley splitting of the bottom conduction band.

In conclusion, through first-principles analysis, we have shown that among all dynamically stable magnetic MA$_{2}$Z$_{4}$ monolayers, only four magnetic nitrides are thermodynamically stable and this occurs under some experimentally challenging growth conditions. In the group of 2D magnetic nitrides, VSi$_2$N$_4$ has been found to be the most stable monolayer, from which two  Janus ferromagnetic monolayers, namely, VSiGeN$_{4}$ and VSiSnN$_{4}$, have been designed. The dynamical stability and thermal stability of these Janus monolayers ensure their non equilibrium growth. Alternatively, a thermodynamically stable VSiGeN$_4$ monolayer can be synthesized in  equilibrium conditions. The electronic structures reveal that they are narrow-band-gap semiconductors. However, the most exciting feature of VSiGeN$_{4}$ and VSiSnN$_{4}$ monolayers is their  weak easy-plane magnetic anisotropy, which is primarily associated with the MSA and the contrasting semiconducting nature (direct vs indirect). The  magnetic properties of these 2D ternary layered vanadium-based Janus semiconductors make them emerging candidates for spintronics and optoelectronics applications.  Besides, the possibility of realization of skyrmions in VSiGeN$_{4}$ and VSiSnN$_{4}$ monolayers can also be investigated, as it was for other Janus monolayers as reported in the contemporary literature~\cite{sky-Liang,sky-Xu}. These results will provide  guidance for  experimental as well as theoretical explorations of the magnetic monolayers of the MA$_2$Z$_4$ family and these two  Janus VSiXN$_4$ monolayers.

This work has been supported by the U.S. DOE, Office of Science, Office of Basic Energy Sciences, under Award No. DE-SC0021127. D.D. thanks Harrison LaBollita for the useful discussion on MSA calculations. D.D. and L.Y. gratefully acknowledge the assistance of the Advanced Computing Group of the University of Maine System for providing computational resources for this work.

\bibliography{reference_final.bib}

\end{document}


\section{\label{}S1. Computational Methods}
Density functional theory (DFT) calculations have been performed using a plane-wave basis with a kinetic energy cutoff of 500 eV and the projector augmented-wave (PAW) method\cite{PAW_bloch} as implemented in the Vienna Ab initio Simulation Package (VASP).~\cite{vasp1,vasp2} The Perdew-Burke-Ernzerhof (PBE)\cite{GGA_pbe} version of the generalized gradient approximation (GGA) has been used as exchange-correlation functional. Both in-plane lattice constants and atomic positions are relaxed using spin-polarized GGA+$U$ until the total energy is converged to 10$^{-8}$ eV, and the forces on each atom are converged to 0.001 eV/\AA. The reciprocal space integration has been carried out using a 15$\times$15$\times$1 Monkhorst-Pack k-mesh grid. The phonon dispersions are calculated using density functional perturbation theory (DFPT)~\cite{dfpt} as implemented in the PHONOPY code \cite{Phonopy} with a 4$\times$4$\times$1 supercell and 3$\times$3$\times$1 k-mesh. Ab initio molecular dynamics simulations are conducted on a 4$\times$4$\times$1 supercell by employing a canonical ensemble and the Nos\'{e} Hoover method.~\cite{MD}

\section{\label{}S2. Thermodynamic stability of MA$_2$N$_4$ monolayers}
 \begin{figure}[H]
\centering \includegraphics[width=0.7\textwidth,clip]{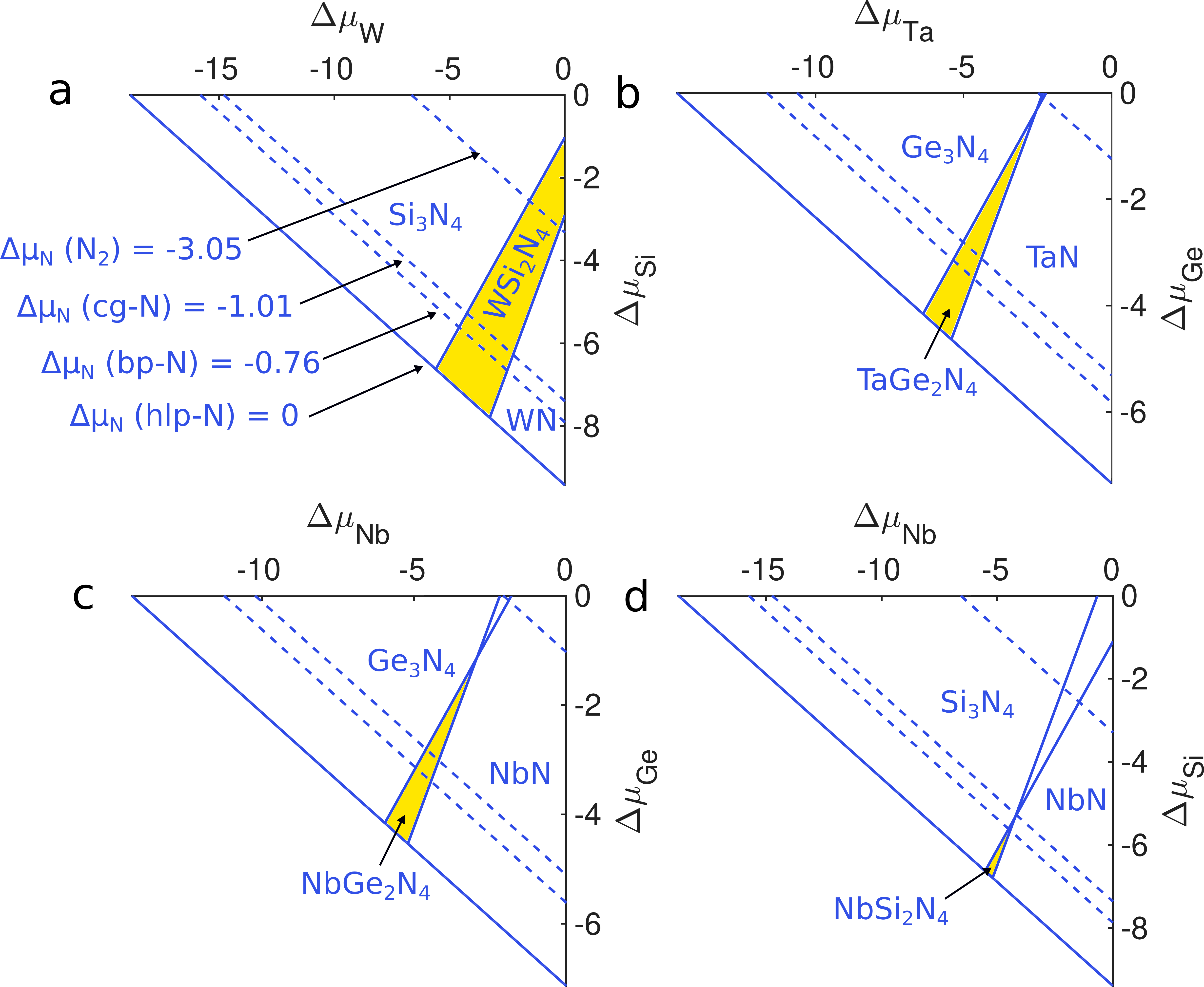}
\caption{2D chemical potential regions (shaded in yellow) where (a) WSi$_2$N$_4$, (b) TaGe$_2$N$_4$, (c) NbGe$_2$N$_4$, and (d) NbSi$_2$N$_4$ monolayers are thermodynamically stable against the formation of constituent elements and their respective competing binary compounds. The shaded area behind the solid (dotted) blue line in the phase diagram is the chemical potential region where these monolayers can be synthesized using chemically rich (poor) N. All the energies are in eV.}
\end{figure}

\section{\label{}S3. Estimation of $U$ for Janus monolayers}
\begin{figure}[H]
\centering \includegraphics[width=0.6\textwidth,clip]{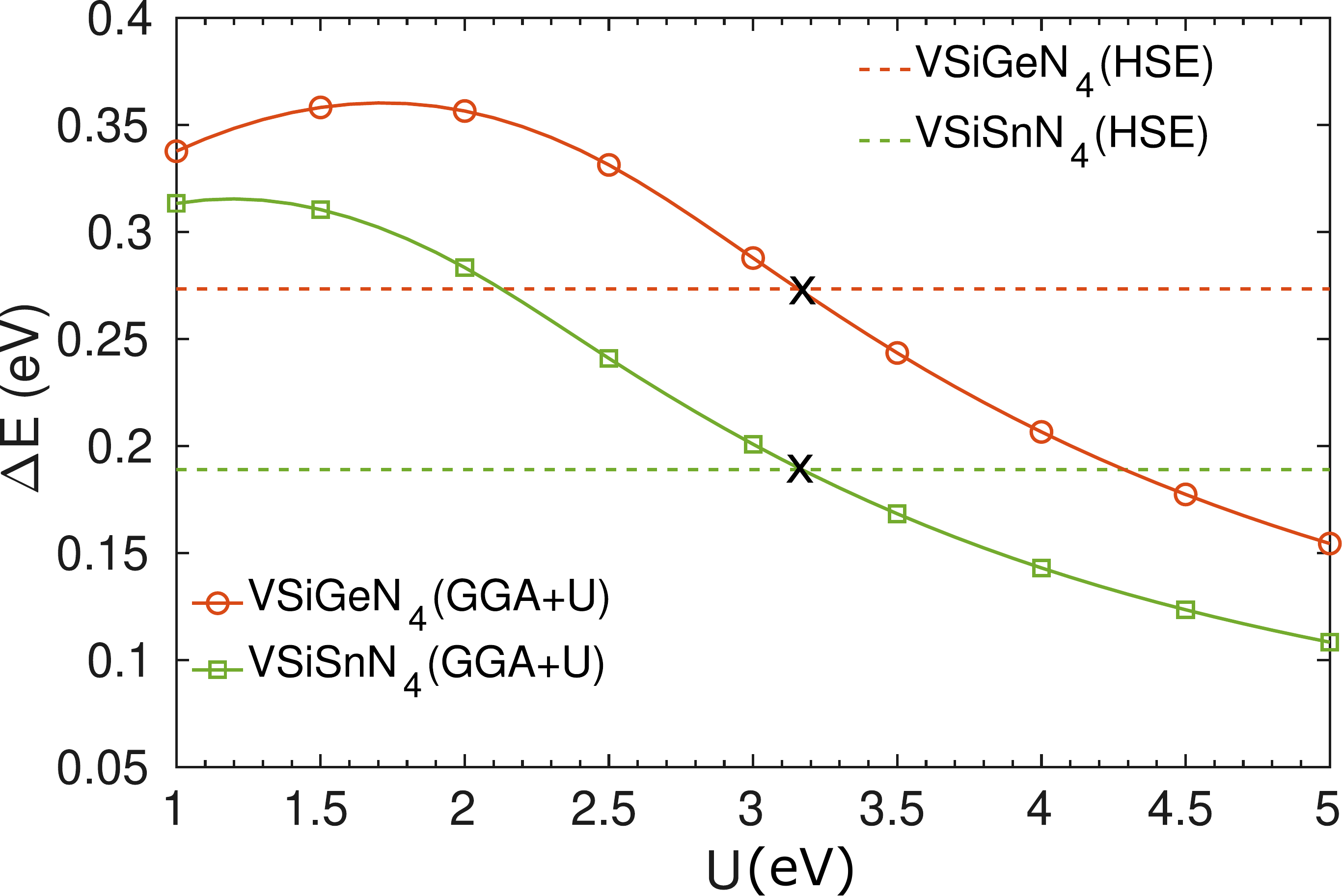}
\caption{Energy difference $\Delta$E = E$_{AFM}$-E$_{FM}$ between AFM and FM spin configurations, as a function of $U$ calculated using the GGA+$U$ methods ($\Delta$E$_{+U}$). The red and green horizontal dotted lines label the $\Delta$E$_{HSE}$ calculated using the HSE06 functional for VSiGeN$_4$ and VSiSnN$_4$ monolayers, respectively. The points where $\Delta$E$_{+U}$ and  $\Delta$E$_{HSE}$ become equal are marked as X.}
\end{figure}

\section{\label{}S4. Exchange constant and transition temperature calculations details}
The magnetic transition temperature T$_{BKT} =$ 1.335 $J/K_B$ is obtained from the results of the Monte Carlo simulation for 2D-$XY$ magnets on a triangular lattice~\cite{BKT_Zhang_triangular}. Here, k$_{B}$ is the Boltzmann constant, and J is the exchange coupling parameter calculated by mapping the DFT total energy to the Heisenberg spin model. In a FM configuration, each V has six neighbors with the same spin (Fig. 4(b) of the manuscript). But in AFM configuration, V four neighbors have opposite spins, and two have the same spin (Fig. 4(d) of the manuscript). Hence, J has been calculated directly from the energy difference $\Delta$E$_m$ = 8JS$^2$ (where S=$\frac{1}{2}$) between AFM and FM spin configurations of each monolayer (Table. 1 of the manuscript). We estimate T$_{BKT}$ = 687 K for VSi$_2$N$_4$, 507 K for VSiGeN$_4$, and 347 K for VSiSnN$_4$ monolayers, respectively.

\section{\label{}S5. Magnetic shape anisotropy (MSA) calculations}
\begin{figure}[hbt!]
\centering \includegraphics[width=.6\textwidth,clip]{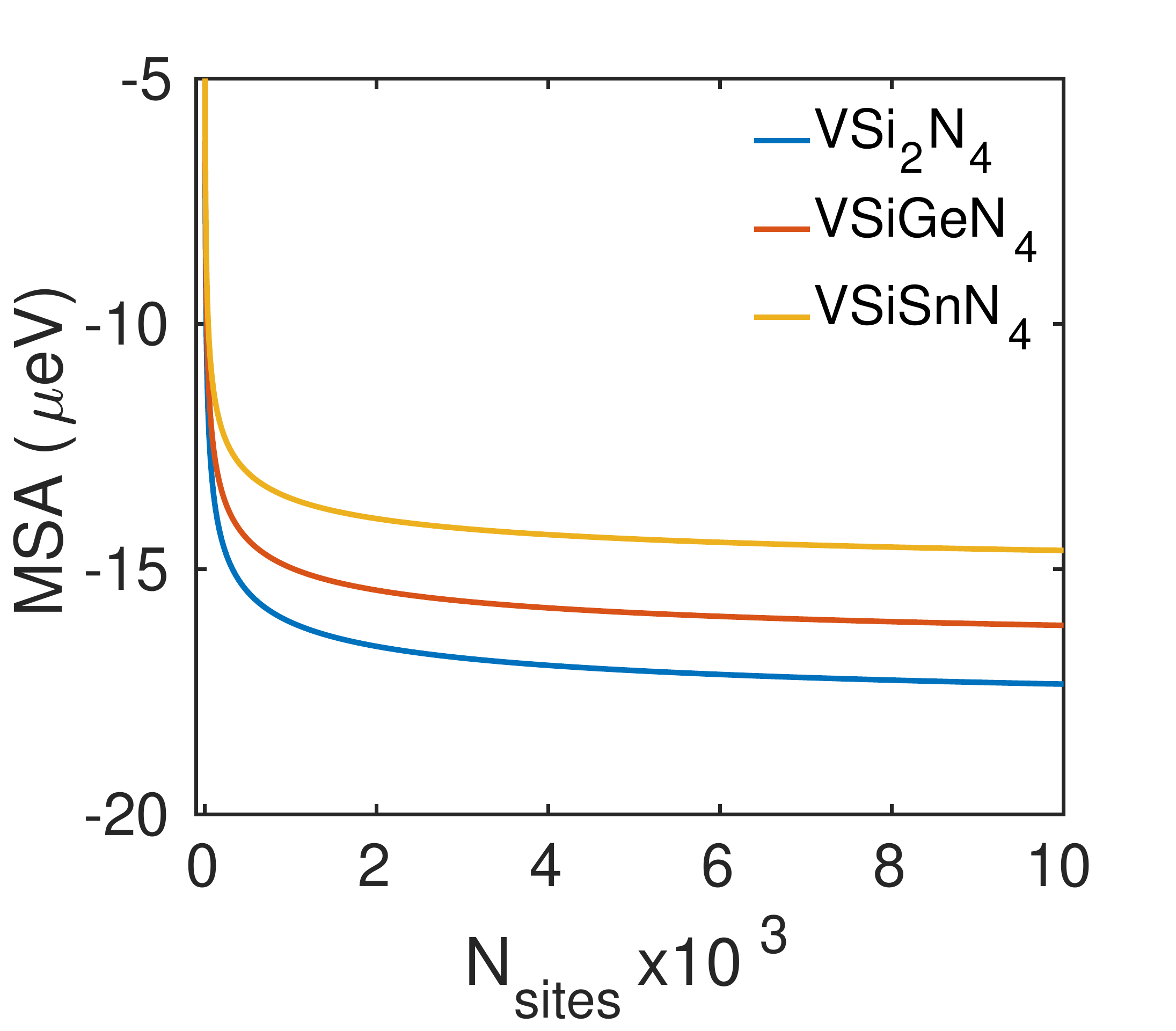}
\caption{Convergence of the MSA energy with respect to the size of the lattice.}
\label{fig3}
\end{figure}

Magnetic anisotropy energy (MAE) measures the dependence of the energy on the orientation of magnetization. Magnetic anisotropy originates mainly from two sources. (i) Magnetocrystalline anisotropy (MCA), which is an intrinsic property of the material caused by the spin-orbit coupling (SOC). (ii) Magnetic shape anisotropy (MSA), which is basically the anisotropic dipolar interaction (Eq.~\ref{eqn1}) of free magnetic poles that tend to magnetize magnetic elements with magnetic moments directed parallel to the surfaces (to minimize magnetostatic energy). Usually, the MAE can be characterized by the MCA. However, for materials with weak SOC, MSA becomes relevant and contributes to MAE. We have calculated MSA energies~\cite{Akram_NL_CrX3} (E$_{MSA}=$ E$_x^{d-d}$ $-$ E$_z^{d-d}$) for VSi$_2$N$_4$, VSiGeN$_4$, and VSiSnN$_4$ monolayers using Eq.~\ref{eqn1}. The values are listed in Table~\ref{table:1} along with MCA and MAE values. 

\begin{equation}
E^{d-d}= \frac{1}{2}\frac{\mu_0}{4\pi}\sum_{i \neq j}\frac{1}{r_{ij}^3}\Big[(\mathbf{m}_i \cdot \mathbf{m}_j)-3\frac{(\mathbf{m}_i \cdot \mathbf{r}_{ij})(\mathbf{m}_j\cdot \mathbf{r}_{ij})}{r_{ij}^2}\Big], 
\label{eqn1}
\end{equation}
Here {\bf m}$_i$ represents the local magnetic moments and {\bf r}$_{ij}$ are vectors that connect the sites i and j. MSA has been calculated as a function of the size of the lattice until the convergence is achieved (Fig.~\ref{fig3}).

\begin{table}
\centering
\begin{center}
\begin{tabular}{p{2cm} p{1.4cm} p{1.4cm} p{1.4cm} p{0.01cm}}
\hline 

\centering Monolayer &\centering MCA & \centering MSA & \centering MAE & \\
 &  \centering ($\mu$eV) & \centering ($\mu$eV) & \centering ($\mu$eV) & \\
 \hline
 \centering VSi$_2$N$_{4}$  & \centering $-$42 & \centering $-$17 & \centering $-$59 & \\

 \hline
 \centering VSiGeN$_{4}$  & \centering $-$4 & \centering $-$16 & \centering $-$20 & \\
 \hline
 \centering VSiSnN$_{4}$ & \centering $+$4 & \centering $-$14 & \centering $-$10 & \\

\hline         
\hline 
\end{tabular}
\caption{Magnetic anisotropy energies per V-site as obtained from DFT calculations. Negative values indicate easy-plane anisotropy for spins.}
\label{table:1}
\end{center}
\end{table}

\section{\label{}S6. Electronic band structures of VSi$_2$N$_4$ and VSiXN$_4$ monolayers within GGA}
\begin{figure}[hbt!]
\centering \includegraphics[width=.95\textwidth,clip]{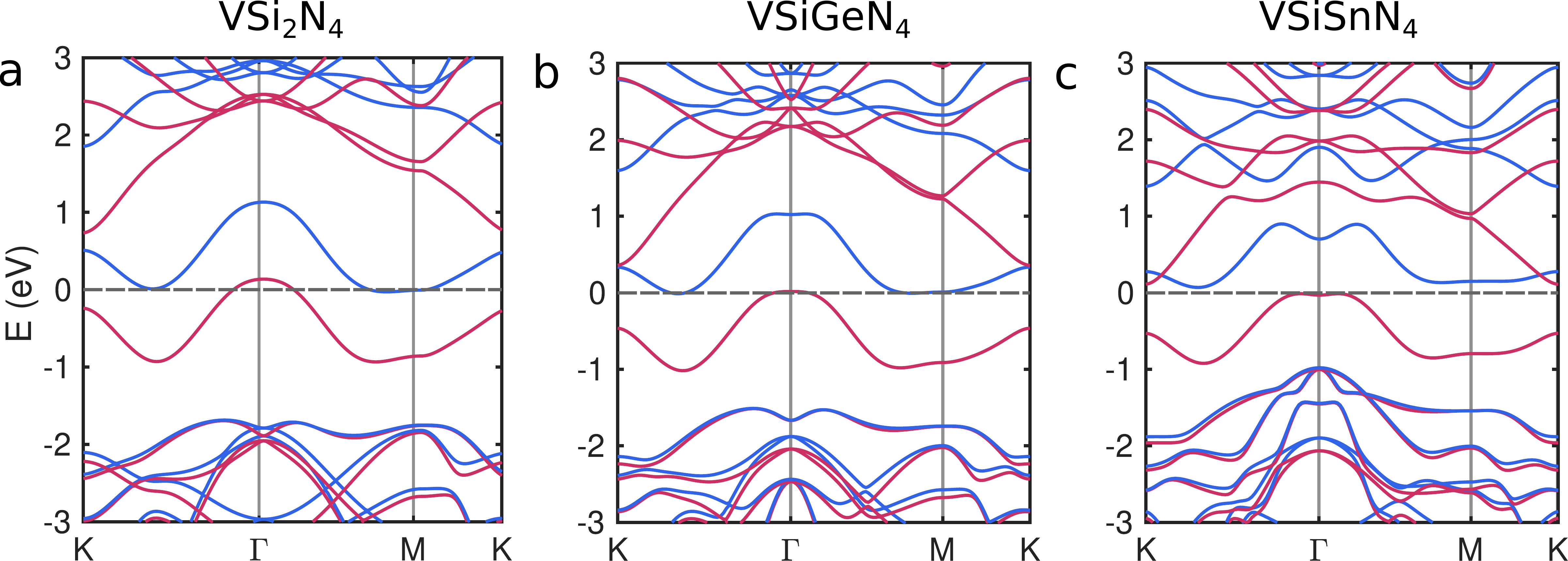}
\caption{GGA band structures of (a) VSi$_2$N$_4$, (b) VSiGeN$_4$, and (c) VSiSnN$_4$. Red and blue lines in the band structures represent the majority and minority spin channels, respectively.}
\end{figure}

\section{\label{}S7. Electronic band structures of VSi$_2$N$_4$ and VSiXN$_4$ monolayers within GGA+$U$}
\begin{figure}[H]
\centering \includegraphics[width=0.95\textwidth,clip]{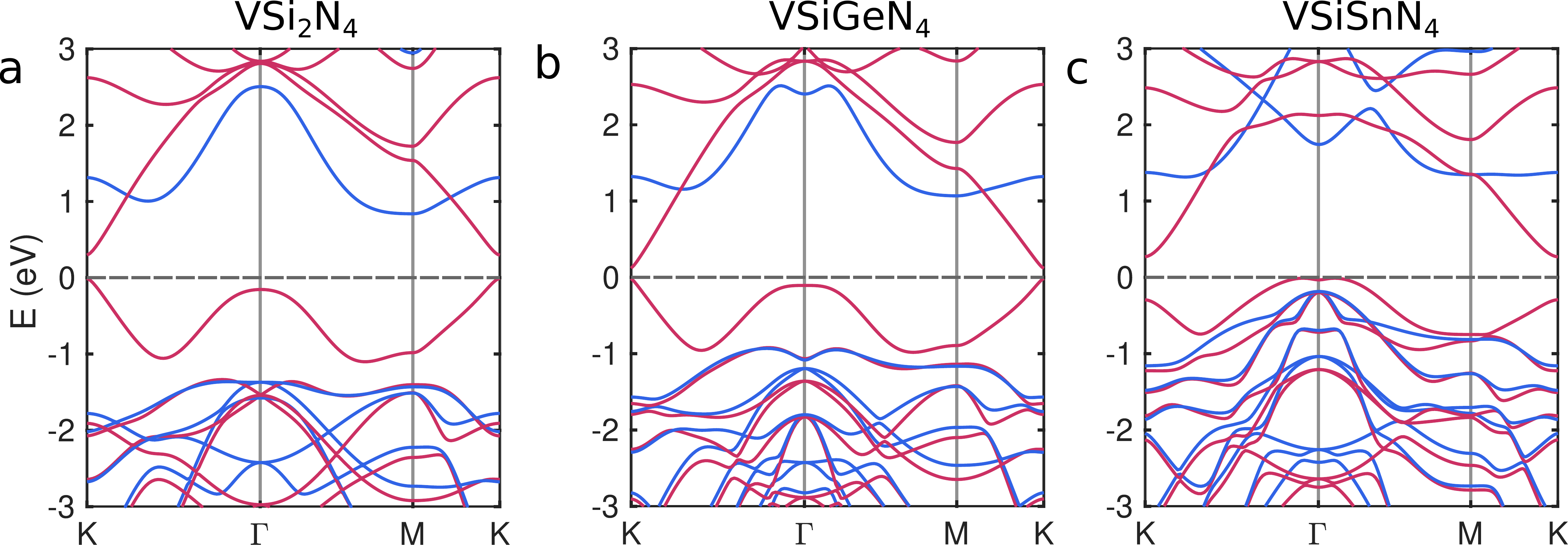}
\caption{GGA+$U$ band structures of (a) VSi$_2$N$_4$, (b) VSiGeN$_4$, and (c) VSiSnN$_4$ monolayers. Red and blue lines in the band structures represent the majority and minority spin channels, respectively.}
\end{figure}

\section{\label{}S8. Electronic band structures of VSi$_2$N$_4$ and VSiXN$_4$ monolayers within GGA+$U$+SOC}
\begin{figure}[H]
\centering \includegraphics[width=0.9\textwidth,clip]{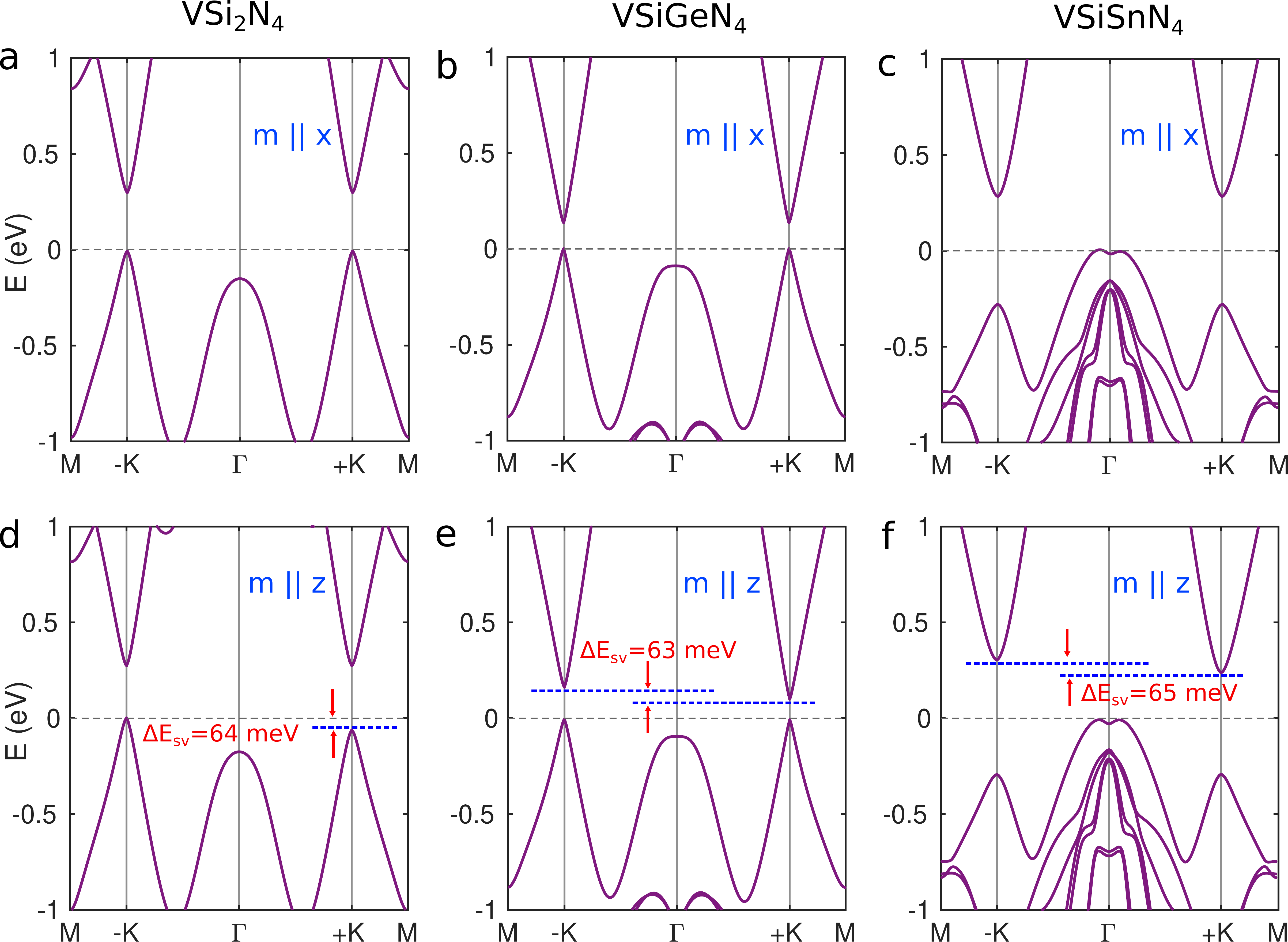}
\caption{GGA+$U$+SOC band structures of (a, d) VSi$_2$N$_4$, (b, e) VSiGeN$_4$, and (c, f) VSiSnN$_4$ monolayers along the M-($-$K)-$\Gamma$-(+K)-M path for the magnetization direction along +x (top panel) and +z (bottom panel) directions.}
\end{figure}

\bibliography{suppl.bib}